\documentstyle[12pt]{article}

\title{A Family of unitary higher order equations
\thanks{\it{This work was partially supported by Consejo Nacional de
Investigaciones Cientificas and Comision de Investigaciones
Cientificas de la Pcia. de Buenos Aires; Argentina.}}}
\author{C.G.Bollini \\
Departamento de Fisica, Fac. de Ciencias Exactas,\\
Universidad Nacional de La Plata.\\
C.C. 67 (1900) La Plata. Argentina. \\
L.E.Oxman \\
Departamento de Fisica, Fac. de Ciencias Exactas, \\
Universidad Nacional de Buenos Aires. \\
Ciudad Universitaria, 1428, \\
Buenos Aires. Argentina. \\
M.C.Rocca \\
Departamento de Fisica, Fac. de Ciencias Exactas, \\
Universidad Nacional de La Plata. \\
C.C. 67 (1900) La Plata. Argentina. \\
and \\
Departamento de Matematicas, Fac. de Ciencias Exactas \\
Universidad Nacional del Centro de la Pcia de Bs. As.\\
Pinto 390, C.P.7000 , Tandil. Argentina.}

\date{September 22, 1995}

\begin{document}

\maketitle

\pagebreak

\begin{abstract}
A scalar field obeying a Lorentz invariant higher order wave equation,
is minimally coupled to the electromagnetic field. The propagator and
vertex factors for the Feynman diagrams, are determined. As an example
we write down the matrix element for the Compton effect. This matrix
element is algebraically reduced to the usual one for a charged
Klein-Gordon particle. It is proved that the $ n^{th} $ order theory
is equivalent to n independent second order theories. It is also
shown that the higher order theory is both renormalizable and unitary
for arbitrary n.


\end{abstract}

\eject

\section{Introduction}

In a previous work (ref.\cite{tp1}), we considered the interaction
of tachyons with the electromagnetic field. As the former can not
exist in free particle states (ref.\cite{tp2},\cite{tp3}), we took
a fourth order wave equation implying two modes of propagation for
a scalar field $ \varphi $ . One of the two modes corresponds to a
normal Klein-Gordon particle. The other is a tachyon mode (ref.
\cite{tp4}). When $ \varphi $ is coupled to electromagnetism by
using the gauge covariant derivative:

\begin{equation}
{\partial}_{\mu} \rightarrow {{\cal D}}_{\mu}= {\partial}_{\mu} -
ieA_{\mu}
\end{equation}

we found that all matrix elements of the fourth order theory:

\begin{equation}
\left( {\Box}^{'2} - m^4 \right) \varphi = 0
\end{equation}
\begin{equation}
{\Box}^{'2} = {{\cal D}}^{\mu} {{\cal D}}_{\mu}
\end{equation}

are equivalent to a second order theory in which the bradyon
(a bradyon is a particle whose momentum $ P_{\mu}$ satisfies
$ P_{\mu} P^{\mu} = -m^2 $) and
the tachyon are independent fields obeying:

\begin{equation}
\left( {\Box}^{'} - m^2 \right) {\varphi}_1 = 0
\end{equation}
\begin{equation}
\left( {\Box}^{'} + m^2 \right) {\varphi}_2 = 0
\end{equation}

The bradyon mode is equivalent to a normal charged Klein-Gordon
particle. The tachyon mode can only be found in closed loops
connected to photon lines.

The fourth order equation belongs to a family found in reference
\cite{tp5}, when studing supersymmetry in spaces of arbitrary
dimensions. In this work the fields obey higher order equations
of motion. The order increasing with de dimensionality of the
space-time. Those equations have the form:

\begin{equation}
\left( {\Box}^n - m^{2n} \right) \varphi = 0
\end{equation}

The usual Klein-Gordon equation is a member of the family $ (n=1) $.
For $ n=2 $ we have the equation examined in reference \cite{tp4} .
On the other hand we have a theory such as Quantum Gravity, where
a perturbative calculation leads to non-renormalizable divergences
proportional to powers of the curvature tensor. There are cases
in which starting with terms cuadratic in the curvature, one
obtains divergences that can be removed by renormalization. These
theories also give rise to fourth order equations.

Equation (6) implies n modes of propagation for the field
$ \varphi $ (ref. \cite{tp6}). It can also be written as:

\begin{equation}
\prod\limits_{s=1}^n \left( \Box - e_s m^2 \right) \varphi = 0
\end{equation}

where

\begin{equation}
e_s = e^{ \frac {2 \pi i} {n} \left( s-1 \right) } \;\;\;\;
\left( s=1,2,...,n \right)
\end{equation}

Of course, $ e_1 = 1 $ for arbitrary n, so that in (7) we always
have a Klein-Gordon factor.

For n=2l (l integer), $ e_{l+1} = -1 $ and we have a tachyon mode.
For n=odd number no tachyon appears. The only real mass is otained
for s=1. All other masses come in complex conjugate pairs.

Except for the s=1 state which is a normal bradyonic mode whose
propagator is Feynman's causal function, none of the states with
$ s \neq 1 $ can propagate as asymptotically free waves. The
corresponding propagators are half advanced and half retarded
(ref.\cite{tp7}). This type of Green function was used in ref.
\cite{tp8} to describe the electromagnetic interaction of
``perfect absorbers'' I.e. when no asymptotic wave escapes the system.

In the next paragraphs we will analyze the behaviour of
$ \varphi $, when the electromagnetic field is introduced in
eq.(6) by means of the gauge covariant derivative eq.(1). I.e.:

\begin{equation}
\Box \rightarrow {\Box}^{'} \equiv \Box - 2ieA \cdot \partial -
e^2 A^2 \;\;,\;\; \left( {\partial}_{\mu} A^{\mu} = 0 \right)
\end{equation}

With this substitution , eq.(6) is transformed into:

\begin{equation}
\left( {\Box}^{'n} - m^{2n} \right) \varphi = 0
\end{equation}

and of course, eq.(7) changes to:

\begin{equation}
\prod\limits_{s=1}^n \left( {\Box}^{'} - e_s m^2 \right) \varphi
= 0
\end{equation}

\section{Interaction terms and propagators}
When we use the substitution given by eq.(9), the iterated D'Alambertian
$ {\Box}^n $ gives rise to the interaction terms of order n:

\begin{equation}
{\Box}^n \rightarrow {\Box}^{'n} = {\left( \Box - 2ieA \cdot \partial -
e^2 A^2 \right)}^n
\end{equation}

The development of (12) gives a polynomial in e of degree 2n:

\begin{equation}
{\Box}^{'n} = {\Box}^n + e {\tilde{I}}_1^{\left( n \right)} + e^2
{\tilde{I}}_2^{\left( n \right)} + \cdot \cdot \cdot + e^{2n}
{\tilde{I}}_{2n}^{\left( n \right) }
\end{equation}

We are going to find the first few terms of (13). I will then be easy
to guess the form of any term.

The first order term can only come from terms containing one factor
$ A \cdot \partial $ and (n-1) powers of the D'Alambertian.

\[ {\tilde{I}}_1^{\left( n \right) } = -2i \left( {\Box}^{n-1}
A \cdot \partial + {\Box}^{n-2} A \cdot \partial \Box +...+
\right.\]
\begin{equation}
\left. + \Box A \cdot \partial {\Box}^{n-2} + A \cdot \partial
{\Box}^{n-1} \right)
\end{equation}

When we take the Fourier transform of (14), the derivative operator
$ (-i {\partial}_{\mu}) $ is transformed into the momentum vector
$ p_{\mu} $ . The D'Alambertian is transformed into $ p^2 $ . The
vector $ A_{\mu} $ leaves its place to the polarization vector of the
photon $ {\epsilon}_{\mu} $ .

\[ I_1^{\left( n \right)} = {\left( -1 \right)}^{n-1} 2 \left(
p^{2 \left( n-1 \right)} \epsilon \cdot p + p^{2 \left( n-2 \right)}
\epsilon \cdot p q^2 + \cdot \cdot \cdot + \epsilon \cdot p
q^{2 \left( n-1 \right)}\right)\]
\begin{equation}
I_1^{\left( n \right)} = 2 {\left( -1 \right)}^{n-1} \epsilon \cdot p
P^{n-1} \left( p^2,q^2 \right); \;\;\;q=p+k \;\;\;\epsilon \cdot p
= \epsilon \cdot q
\end{equation}
\begin{equation}
P^t \left( p^2,q^2 \right)= \sum\limits_{s=0}^{s=t} p^{2 \left(t-s
\right)} q^{2s}
\end{equation}

>From now on we will write all interaction terms in momentum space.
The second order term in e contains a part in $ A^2 $ (cf. eq.(12)),
which is similar to (15) an another part in $ {\Box}^a $
$ A \cdot \partial $ $ {\Box}^b $ $ A \cdot \partial $ $ {\Box}^c $,
with a+b+c=n-2.

\[ I_2^{\left( n \right)} = 2 {\left( -1 \right)}^{n-1} {\epsilon}_1
\cdot {\epsilon}_2 P^{n-1} \left( p^2,q^2 \right)+ 4 {\left( -1
\right)}^{n-1}
{\epsilon}_1 \cdot p_1 {\epsilon}_2 \cdot q  \cdot \]
\[ \cdot P^{n-2} \left( p_1^2,q^2,p_2^2 \right) +
4 {\left( -1 \right)}^{n-1} {\epsilon}_2 \cdot p_1 {\epsilon}_1
\cdot p P^{n-2} \left( p_1^2,r^2,p_2^2 \right). \]
\begin{equation}
\left(p_2=p_1+k_1+k_2,\;q=p_1+k_1,\;r=p_1+k_2 \right)
\end{equation}

where

\begin{equation}
P^t \left( p^2,q^2,r^2 \right) = \sum\limits_{a+b+c=t} p^{2a}
q^{2b} r^{2c}
\end{equation}

For the third order we have to pick-up from (12), all terms containig
three factors $ A \cdot \partial $ or one factor $ A \cdot \partial $
and a factor $ A^2 $. The corresponding D'Alembertians (as in (14))
give rise to the P-coefficients whose main properties are going to
be specified in the next paragraph.The other interaction terms are
found in a similar way.

The propagator for eq.(10) can be defined by:

\begin{equation}
\left( {\Box}^n - m^{2n} \right) {\tilde{G}}^{\left( n \right)}=
i \delta
\end{equation}

By Fourier transforming eq.(19) we get:

\begin{equation}
G^{\left( n \right)} = \frac { {\left( -1 \right)}^n i } { p^{2n}
- {\left( -m^2 \right)}^n }
\end{equation}

We now use the identity:

\begin{equation}
\frac {1} { x^n - a^n } = \frac {1} { n a^{n-1}} \sum\limits_{s=1}^n
\frac {e_s} { x - e_s a}
\end{equation}

where $ e_s $ is given by eq.(8).

With $ x=p^2 $ and $ a=-m^2 $, we get:

\begin{equation}
G^{\left( n \right)} = \frac { -i} {n m^{2\left( n-1 \right)}}
\sum\limits_{s=1}^n \frac {e_s} {p^2 + e_s m^2}
\end{equation}

The first term of (22) (s=1) represents the Klein-Gordon propagator.
The other terms correspond to the other modes of propagation. The
common factor $ {(n m^{2(n-1)})}^{-1} $ is the relative normalization
of the wave function whose propagator is defined by (19), with
respect to that of the usual second order equation. To obtain an
n-independent normalization we have to divide each external line
by the factor $ {(n m^{2(n-1)})}^{1/2} $.

The interaction resulting from (12) seems to be of the unrenormalizable
type for $ n>1 $. Compare for example $ I_1^{(n)} $ (eq.(15) for
$ n>1 $ with $ I_1^{(1)} $= 2 $ \epsilon \cdot p $. However, the
propagator (20) has (n-1)  extra powers of $ p^2 $ in the
denominator. So that, by power counting the theory turns out to be
renormalizable. Furthermore, we are going to show that it is
equivalent, for arbitrary n, to the usual Klein-Gordon theory for
a charged scalar particle.

\section{Vertex Factors}

To determine the factors $ P^t $ we will now describe their properties.

Each $ P^t(x_1,...,x_s) $ is a sum over all monomials of degree t,
formed with products of powers of its arguments.

\begin{equation}
P^t \left(x_1,...,x_s\right) = \sum\limits_{a_1+...+a_s=t} x_1^{a_1}
x_2^{a_2}...x_s^{a_s}
\end{equation}
\[ a_l \geq 0 \left( l=1,2,...,s \right) \]

We define

\begin{equation}
P^t=0 \;\; for\;\;t<0 \;\;and \;\;P^0=1
\end{equation}

>From (23) we have

\begin{equation}
P^1 \left( x_1,...,x_s \right) = \sum\limits_{l=1}^s x_l
\end{equation}
\begin{equation}
P^t \left( x \right) = x^t
\end{equation}
\begin{equation}
P^t \left( x,y \right) = \sum\limits_{l=0}^t x^{t-l} y^l
\end{equation}

All $ P^t $ are symmetrical homogeneous functions of their arguments:

\begin{equation}
P^t \left( \alpha x_1, \alpha x_2,..., \alpha x_s \right) =
{\alpha}^t P^t \left( x_1,...,x_s \right)
\end{equation}

We can also write (23) in the form:

\[ P^t \left( x_1,...,x_s \right) = \sum\limits_{a_1=0}^t x_1^{a_1}
\sum\limits_{a_2+...+a_s = t-a_1 } x_2^{a_2} ...x_s^{a_s} \]
\begin{equation}
P^t \left( x_1,...,x_s \right) = \sum\limits_{a_1=0}^t x_1^{a_1}
P^{t-a_1} \left( x_2,...,x_s \right)
\end{equation}

So that:

\[ x_1 P^t \left( x_1,...,x_s \right) = \sum\limits_{a_1=0}^t
x_1^{a_1 + 1} P^{t- a_1} \left( x_2,...,x_s \right) = \]
\[ = \sum\limits_{b=1}^{t+1} x_1^b P^{t+1-b} \left( x_2,...,
x_s \right) = \sum\limits_{b=0}^{t+1} x_1^b P^{t+1-b} \left( x_2,...
,x_s \right) - P^{t+1} \left( x_2,...,x_s \right) \]
\begin{equation}
x_1 P^t \left( x_1,...,x_s \right) = P^{t+1} \left( x_1,x_2,...
x_s \right) - P^{t+1} \left( x_2,...,x_s \right)
\end{equation}

Also:

\[ x_2 P^t \left( x_1,...,x_s \right) = P^{t+1} \left( x_1,x_2,...,
x_s \right) - P^{t+1} \left( x_1,x_3,...,x_s \right) \]

Then

\begin{equation}
\left( x_1 - x_2 \right) P^t \left( x_1,...,x_s \right) =
P^{t+1} \left( x_1,x_3,...,x_s \right) - P^{t+1} \left( x_2,x_3,
...,x_s \right)
\end{equation}

In particular

\begin{equation}
\left( x_1 - x_2 \right) P^t \left( x_1,x_2 \right) = P^{t+1} \left(
x_1 \right) - P^{t+1} \left( x_2 \right) = x_1^{t+1} - x_2^{t+1}
\end{equation}

If we choose $ x_1 = p^2 $ and $ x_2 = -m^2 $ , we get from (32) :

\begin{equation}
\left( p^2 + m^2 \right) P^{n-1} \left( p^2, -m^2 \right) =
p^{2n} - {\left( -m^2\right)}^n
\end{equation}

And the denominator of the Green function (20), factorizes according
to (33):

\begin{equation}
G^{\left( n \right) } = \frac { {\left( -1 \right)}^n i } { \left(
p^2 + m^2 \right) P^{n-1} \left( p_1^2 - m^2 \right) }
\end{equation}

\section{Compton effect}

We are now ready to evaluate the cross section for any physical
process, in a given perturbative order, for any higher order
equation of the family (10).

We will take as an example the second order Compton effect. The
initial and final momentum of the charged bradyon are $ p_1 $
and $ p_2 $. The incoming photon has a polarization $ {\epsilon}_1 $
and a momentum $ k_1 $. The final photon has a polarization
$ {\epsilon}_2 $ and momentum $ k_2 $. We define $ p= $ $ p_1
+ k_1 = $ $ p_2 + k_2 $ , $ q= $ $ p_1-k_2= $ $ p_2-k_1 $ ;
$ {\epsilon}_1 \cdot k_1 $ $ =0 $ $ {\epsilon}_2 \cdot k_2 $ $ =0 $.

The matrix element corresponding to eq.(10), with the propagator (20)
and the interaction vertices (15) and (17), is:

\[ M^{\left( n \right) } =\;\;\;\;\;\;\;\;\;\;\;\;\;\;\;\;\;\;\;\;
\;\;\;\;\;\;\;\;\;\;\;\;\;\;\;\;\;\;\;\;\;\;\;\;\;\;\;\;\;\;\;\;\; \]
\[\left[ 2i {\left(-1\right)}^{n-1}
{\epsilon}_1 \cdot p_1 P^{n-1} \left( p_1^2,p^2 \right) \right]
\frac { {\left(-1\right) }^n i } { p^{2n} - {\left( -m^2 \right)}^n }
\left[ 2i {\left(-1\right)}^{n-1} {\epsilon}_2 \cdot p_2 P^{n-1}
\left( p^2,p_2^2 \right) \right]+\]
\[+ \left[ 2i {\left(-1\right)}^{n-1}
{\epsilon}_2 \cdot p_1 P^{n-1} \left(p_1^2,q^2 \right) \right]
\frac { {\left(-1\right)}^n i } { q^{2n} - {\left(-m^2 \right)}^n }
\left[ 2i{\left(-1\right)}^n {\epsilon}_1 \cdot p_2 P^{n-1} \left(
q^2,p_2^2 \right) \right] + \]
\[ i {\left( -1\right)}^n \lbrace 4 {\epsilon}_1 \cdot p_1
{\epsilon}_2 \cdot p_2 P^{n-2} \left(p_1^2,p^2,p_2^2\right) + 4
{\epsilon}_2 \cdot p_1 {\epsilon}_1 \cdot p_2 \cdot \]
\[\cdot P^{n-2} \left(p_1^2,q^2,p_2^2\right) + 2{\epsilon}_1
\cdot {\epsilon}_2 P^{n-1} \left(p_1^2,p_2^2 \right) \rbrace \]
\[ M^{\left(n\right)}= 4i{\left(-1\right)}^n \left\{ {{\epsilon}_1
\cdot p_1 {\epsilon}_2 \cdot p_2 \left[ \frac {P^{n-1}\left(p_1^2,
p^2 \right) P^{n-1} \left(p^2,p_2^2 \right) } {p^{2n}- {\left(-m^2
\right)}^n } - P^{n-2} \left(p_1^2,p^2,p_2^2\right) \right] + }\right. \]
\[ \left. {{\epsilon}_2 \cdot p_1 {\epsilon}_1 \cdot p_2 \left[ \frac {
P^{n-1}\left(p_1^2,q^2\right) P^{n-1}\left(q^2,p_2^2\right) }
{q^{2n}-{\left(-m^2\right)}^n } - P^{n-2}\left(p_1^2,q^2,p_2^2
\right) \right] } \right\} + \]
\begin{equation}
+2i{\left(-1\right)}^n {\epsilon}_1 \cdot {\epsilon}_2
P^{n-1} \left(p_1^2,p_2^2 \right)
\end{equation}

In the last equation we use (33) and $ p_1^2=-m^2 $ , $ p_2^2=
-m^2 $ :

\[ M^{\left(n\right)}= 4i {\left(-1\right)}^n \left\{ {{\epsilon}_1
\cdot p_1 {\epsilon}_2 \cdot p_2 \left[ \frac { P^{n-1} \left(-m^2,p^2
\right)} {p^2+m^2} - \frac {\left(p^2+m^2\right) P^{n-2}\left(-m^2,
p^2,-m^2 \right)} {p^2+m^2} \right] +}\right. \]
\[ \left.{{\epsilon}_2 \cdot p_1 {\epsilon}_1 \cdot p_2 \left[
\frac {P^{n-1}\left(-m^2,q^2\right)} {q^2+m^2} - \frac {\left(
q^2+m^2\right) P^{n-2} \left(-m^2,p^2,-m^2\right)} {p^2+m^2} \right]
}\right\} + \]
\begin{equation}
2i{\left(-1\right)}^n {\epsilon}_1 \cdot {\epsilon}_2 P^{n-1}
\left(-m^2,-m^2\right)
\end{equation}

But, according to (31):

\begin{equation}
\left(x+m^2\right)P^{n-2} \left(-m^2,x,-m^2\right) = P^{n-1}
\left(x,-m^2\right) - P^{n-1} \left(-m^2,-m^2\right)
\end{equation}

And, according to (27):

\begin{equation}
P^{n-1} \left(x,x\right)= n x^{n-1}
\end{equation}

So, we finally get, for the normalized matrix element $
{\bar{M}}^{(n)} $

\[{\bar{M}}^{\left(n\right)}= {\left( n m^{2\left(n-1\right)}
\right)}^{-1} M^{\left(n\right)} = \]
\begin{equation}
=-4i \left( {\epsilon}_1 \cdot p_1 {\epsilon}_2 \cdot p_2
\frac {1} {p^2+m^2} + {\epsilon}_2 \cdot p_1 {\epsilon}_1
\cdot p_2 \frac {1} {q^2+m^2} \right) -2i {\epsilon}_1 \cdot
{\epsilon}_2
\end{equation}

We can see from (39) the interesting fact that, no matter how high
the order of the equation (10) is, we always end up with the
matrix element corresponding to the second order Klein-Gordon
equation coupled to the electromagnetic field.

The same fact is true when we consider, for example a ``multiphoton''
scattering, in which a collision of a photon with a charged bradyon
produces any number of scattered photons. We do not intend to give
here a direct proof, which involves some laborious algebraic
manipulation of the matrix elements. In ref.\cite{tp9} we show an
explicit evaluation for the double photon scattering. Anyway, we
are going to present a general proof in \S 5.

We can also answer the following question: what is the amplitude for
the Compton effect to produce a change from the bradyon mode to
any other mode of the field $ \varphi $ (solution of (10))?

To answer this question, we take again the matrix element
$ M^{(n)} $ , eq.(35). But instead of taking $ p_2^2 $ = $ -m^2 $ ,
we use $ p_2^2 $ = $ -e_s m^2 $ ; where $ e_s $ is given by eq.(8).

It is easy to follow the procedure that leads to eqs.(36) to (39).
Now the matrix element is proportional to (cf. eq.(27)) :

\[P^{n-1} \left(-m^2,-e_s m^2 \right) = \sum\limits_{l=0}^{n-1}
{\left(-m^2\right)}^{n-1-l} {\left(-m^2e_s\right)}^l = \]
\[= {\left(-m^2\right)}^{n-1} \sum\limits_{l=0}^{n-1} e_s^l =
{\left(-m^2\right)}^{n-1} \frac {1-e_s^n} {1-e_s} \]

For any s, $ e_s^n =1 $ (cf. eq.(8)). So that when $ e_s \neq 1 $ ,

\begin{equation}
P^{n-1} \left(-m^2,-e_sm^2\right) = 0 \;\;\;,\;\;\;s \neq 1
\end{equation}

Eq. (40) tells us that the probability amplitude for a change from
a bradyon mode to any other (different) mode is exactly zero.

\section{General proof}

We will use functional methods to establish the equivalence of
the general order equation (10) and the second order one.

The Lagrangian corresponding to the field $ \varphi $ interacting
minimally with the electromagnetic field is:

\begin{equation}
{\cal L} = - \frac {1} {4} F_{\mu\nu}F^{\mu\nu} + \bar{\varphi}
\left( {\Box}^{'n} - m^{2n} \right) \varphi
\end{equation}

where $ {\Box}^{'} $ is defined by eq.(19).

Our generating functional is:

\begin{equation}
{\cal Z} \left( {\cal J}, \bar{{\cal J}}, {\cal K} \right) =
\int \left[ {\cal D} A \right] \left[ {\cal D} \varphi \right]
\left[ {\cal D} \bar{\varphi} \right] e^{i \int dx \left(
{\cal L} + {\cal J} \varphi + \bar{{\cal J}} \bar{\varphi} +
{\cal K} A \right) }
\end{equation}

where $ {\cal J} $, $ \bar{{\cal J}} $ and $ {\cal K} $ are external
sources.
To assure the Lorentz gauge for $ A_{\mu} $ , a term $ \xi {(
{\partial}_{\mu} A^{\mu} )}^2 $ should be added to the Lagrangian
(41).

The exponent in (42) is a quadratic function of the scalar field
$ \varphi $ . We can then use the general gaussian formula
\cite{tp10}:

\[ \int \left[ {\cal D} \varphi \right] \left[ {\cal D}
\bar{\varphi} \right] e^{i \int dx \left( \bar{\varphi}
{\cal Q} \varphi + {\cal J} \varphi + \bar{{\cal J}} \bar{\varphi}
\right) } = \]
\begin{equation}
N^{'} {\mid {\cal Q} \mid}^{-2} e^{-i \int dx {\cal J}
{{\cal Q}}^{-1} \bar{{\cal J}} }
\end{equation}

where $ \mid {\cal Q} \mid $ is the functional determinant of the
operator ${\cal Q}$. For our case we introduce the notations:

\begin{equation}
{\cal P} = {\Box}^{'n} - m^{2n}
\end{equation}
\begin{equation}
{{\cal P}}_s = {\Box}^{'} - e_s m^2 \;\;,\;\; e_s= e^{ \frac {
2 \pi i} {n} \left(s-1\right) }\;\; \left( s=1,...,n\right)
\end{equation}

The factorization of eq.(10), eq.(11), means in our notation,
that:

\begin{equation}
{\cal P} = \prod\limits_{s=1}^n {{\cal P}}_s
\end{equation}

Note also that:

\begin{equation}
\left[ {{\cal P}}_a , {{\cal P}}_b \right] =0 \;\;\;,\;\;\;\left(
a,b=1,...,n\right)
\end{equation}

The functional determinant of {\cal P}, also factorizes according
to (46):

\begin{equation}
\mid {\cal P} \mid = \prod\limits_{s=1}^n \mid {{\cal P}}_s \mid
\end{equation}

>From (46) and (47) we get:

\[ {{\cal P}}^{-1} = \prod\limits_{s=1}^n {{\cal P}}^{-1} \]

For the inverse of the {\cal P}-operators to be properly defined,
we must take into account the boundary conditions imposed on the
Green functions. $ {{\cal P}}_s^{-1} $ $ (s\neq1) $ correspond to
the half advanced and half retarded Wheeler's function. However,
for our purpose here, we do not need to be more specific. It
suffices to give our formulae a symbolic character.

>From the identity (21) we deduce:

\[ \frac {1} {{\Box}^{'n} - m^{2n}} = \frac {1} {nm^{2\left(n-1
\right)}} \sum\limits_{s=1}^n \frac {e_s} {{\Box}^{'}-e_sm^2} \]

Or, using the notations (44), (45):

\begin{equation}
{{\cal P}}^{-1}= \frac {1} {nm^{2\left(n-1\right)}}
\sum\limits_{s=1}^n e_s {{\cal P}}_s^{-1}
\end{equation}

With (48) and (49) we obtain:

\[ (43) = N^{'} \prod\limits_{s=1}^n {\mid {{\cal P}}_s \mid}^{-2}
e^{-i \int \frac {dx} {nm^{2\left(n-1\right)}} \sum\limits_{t=1}^n
{\cal J} e_t {{\cal P}}_t^{-1} \bar{{\cal J}} } \]
\begin{equation}
(43) = N^{'} \prod\limits_{s=1}^n {\mid {{\cal P}}_s \mid}^{-2}
e^{-i \int \frac {dx} {nm^{2\left(n-1\right)}} {\cal J} e_s
{{\cal P}}_s^{-1} \bar{{\cal J}} }
\end{equation}

We now introduce n scalar fields $ {\varphi}_s $ $ (s=1,...,n) $
and we use again the functional gaussian formula (43).

\[ \int \left[ {\cal D} {\varphi}_s \right] \left[ {\cal D}
{\bar{\varphi}}_s \right] e^{i \int dx \lbrace {\bar{\varphi}}_s
{\bar{e}}_s {{\cal P}}_s {\varphi}_s + \frac {1} {\sqrt{n}
m^{\left(n-1\right)}} \left( {\cal J} {\varphi}_s + \bar{{\cal J}}
{\bar{\varphi}}_s \right) \rbrace } = \]
\begin{equation}
= N_s {\mid {\cal P}_s \mid}^{-2} e^{-i \int \frac {dx}
{nm^{2\left(n-1\right)}} {\cal J} e_s {\cal P}_s^{-1} \bar{{\cal J}} }
\end{equation}

After a renormalization of the scalar field:

\[ {\varphi}_s \rightarrow \sqrt{n} m^{\left(n-1\right)}
{\varphi}_s \]

we can introduce (51) in (50) to write:

\[ \int \left[ {\cal D} \varphi \right] \left[ {\cal D}
\bar{\varphi} \right] e^{i \int dx \left( \bar{\varphi} {\cal P}
\varphi + {\cal J} \varphi \bar{{\cal J}} \bar{\varphi} \right) } = \]
\[ N \int \prod\limits_s \left[ {\cal D} {\varphi}_s \right]
\left[ {\cal D} {\bar{\varphi}}_s \right] \cdot \]
\begin{equation}
e^{i \int dx \sum\limits_s \left( nm^{2\left(n-1\right)}
{\bar{\varphi}}_s {\bar{e}}_s {{\cal P}}_s {\varphi}_s +
{\cal J} {\varphi}_s + \bar{{\cal J}} {\bar{\varphi}}_s \right)}
\end{equation}

To obtain again the generating functional (42), we multiply both
members of (52) with:

\[ e^{i \int dx \left( - \frac {1} {4} F_{\mu\nu} F^{\mu\nu} +
{\cal K} \cdot A \right)} \]

and perform a functional integration over $ A_{\mu} $ .

\[ {\cal Z} \left( {\cal J}, \bar{{\cal J}}, {\cal K} \right) =
N \int \left[ {\cal D} A \right] \prod\limits_s \left[ {\cal D}
{\varphi}_s \right] \left[ {\cal D} {\bar{\varphi}}_s \right] \]
\begin{equation}
e^{i \int dx \left( \tilde{{\cal L}} + \sum\limits_t \left(
{\cal J} {\varphi}_t + \bar{{\cal J}} {\bar{\varphi}}_t \right) +
{\cal K} A \right) }
\end{equation}

Or,

\begin{equation}
{\cal Z} \left( {\cal J}, \bar{{\cal J}}, {\cal K} \right) =
\tilde{{\cal Z}} \left( {\cal J}, \bar{{\cal J}}, {\cal K}
\right)
\end{equation}

In (53) we have introduced the definition:

\begin{equation}
\tilde{{\cal L}}= - \frac {1} {4} F_{\mu\nu} F^{\mu\nu} +
\sum\limits_{s=1}^n nm^{2\left(n-1\right)} {\bar{\varphi}}_s
{\bar{e}}_s {{\cal P}}_s {\varphi}_s
\end{equation}

The equivalence of the generating functionals $ {\cal Z} $ and
$ \tilde{{\cal Z}} $, expressed by eqs.(53) and (54), implies
the equivalence of the lagrangians (41) and (55).

The lagrangian (41) describes the gauge invariant electromagnetic
interaction of a scalar field obeying a n-th order equation
(cf. eq.(10)). On the other hand, eq.(55) refers to the electromagnetic
interaction of n independent scalar fields obeying second order
equations:

\begin{equation}
{\cal P}_s {\varphi}_s \equiv \left( {\Box}^{'} -e_s m^2 \right)
{\varphi}_s = 0 \;\;,\;\;\left(s=1,...,n\right)
\end{equation}

For s=1, eq.(56) is a normal Klein-Gordon equation for a charged
scalar particle. For $ s \neq 1 $, $ {\varphi}_s $ is a ``virtual
field''. It can only exist associated with closed loops attached
to photon lines.

The equivalence shown by eq.(54) also tells us about the unitarity
of the higher order theory. In fact, if we take any closed loop
corresponding to one of the eqs.(56), with $ s\neq 1 $ , and note
that the propagators are half advanced and half retarded, we see
that the imaginary part of the loop can only come from the imaginary
part of the mass parameter. Then, the absortive part of this diagram
cancels with another similar one for which the internal lines are
related to the complex conjugate mass parameter (See also
ref.\cite{tp11}). In other words, any absortive part coming from
a loop corresponding to $ e_sm^2 $ $ (s\neq 1) $ is exactly canceled
by another contribution coming from a similar loop corresponding to
$ {\bar{e}}_s m^2 $ .

\section{Discussion}

Starting with a n-th order equation of motion for a scalar field
$ \varphi $, we introduce the electromagnetic interaction by means of
the gauge invariant minimal coupling procedure. The field $ \varphi $
has n modes of propagation. Its evolution can be followed perturbatively
with the Feynman's diagrams techniques. The Green function has n
poles corresponding to n modes of propagation. The probability
amplitude and the cross-section for any physical process can be
determined without ambiguities.

We have shown that it is possible to reduce algebraically, each
matrix element corresponding to the n-th order equation, to simple
matrix elements in which the electromagnetic interaction and the
propagation are described by second order equations.

The general proof of \S 5 shows that $ \varphi $, with its n modes of
propagation, behaves like n independent scalar fields $ {\varphi}_s $ ,
each obeying a simple second order equation.

The equivalence also shows the interesting fact that, no matter how
high the order n is, the theory is unitary and renormalizable. Thus,
we have the equivalence of two different points of view. One of
them is the usual theory for a normal charged Klein-Gordon particle.
The other is the theory for a field that obeys a higher order
equation, minimally coupled to electromagnetism. The algebraic
reduction of the matrix elements for the latter theory to those
of the former one, presented in \S 4, appears to be rather mysterious.
The simplification seems to be the result of a fortuitous coincidence.
However, the functional proof ( \S 5 ), sheds light on the nature of
the equivalence.

It is clear that the proof is based on two fundamental properties
of the equation of motion:

a) {\bf Factorabizability} (cf. eqs. (10) and (11)). I.e.:

\[ {\cal P} \varphi =0 \]
\[ {\cal P} = \prod\limits_{s=1}^n {\cal P}_s\;\;,\;\; \left[
{\cal P}_a, {\cal P}_b \right] =0 \]

b) {\bf Separability} (cf. eq. (22)). I.e., the general propagator can
be expressed as a linear combination of individual propagators.

\[ {{\cal P}}^{-1} = \sum\limits_{s=1}^n {\alpha}_s {{\cal P}}_s^{-1} \]

Any operator which is both, factorizable and separable, gives rise to
an equivalence theorem.

As a matter of facts, any arbitrary higher order equation can be
factorized (we take $ c_n = 1 $ ):

\[ {\cal P} \equiv \sum\limits_{t=0}^n c_t {\Box}^t = \
\prod\limits_{s=1}^n \left( \Box - m_s^2 \right) \]
\[ \prod\limits_{s=1}^n {\cal P}_s \]

where the ``masses'' $ m_s^2 $  are the n roots of $ {\cal P} $ =0.
Also the propagator $ {\cal P}^{-1} $ can be expressed as:

\[ {\cal P}^{-1} = \frac {1} { \prod\limits_{s=1}^n \left( \Box -
m_s^2 \right) } = \sum\limits_{s=1}^n \frac {d_s} { \Box - m_s^2} \]

where $ d_s $ (s=1,...,n) are appropriate constants.

The distinctive feature of eq.(10) is the fact that the masses
have the particular form $ m_s^2 $ = $ e_sm^2 $ , where $ m^2> $ 0
and $ e_s $ are phase factors given by eq.(8).

The form of eq.(6) is dictated by its physical origin, based on
supersymmetry in higher dimensions \cite{tp5}.

\pagebreak


\begin{thebibliography}{99}

\bibitem{tp1} D.G.Barci, C.G.Bollini, M.C.Rocca: ``On the coupling
 of tachyons to electromagnetism''. La Plata preprint, April (1995).
\bibitem{tp2} D.G.Barci, C.G.Bollini, M.C.Rocca: Il Nuovo Cimento
 Vol.106 A, N.5, pag.603(1993).
\bibitem{tp3} D.G.Barci, C.G.Bollini, M.C.Rocca: Int. J. of Mod.
 Phys. A Vol.9, N.20, pag.3497(1994).
\bibitem{tp4} D.G.Barci, C.G.Bollini, M.C.Rocca: Int. J. of Mod.
 Phys. A Vol.10, N.12. pag.1737(1995).
\bibitem{tp5} C.G.Bollini, J.J.Giambiagi: Phys. Rev D Vol.32,
 pag.3316(1985).
\bibitem{tp6} D.G.Barci, C.G.Bollini, L.E.Oxman, M.C.Rocca: Int.
 J. of Mod. Phys. A Vol.9, N.23, pag.4169(1994).
\bibitem{tp7} C.G.Bollini, L.E.Oxman: Int. J. of Mod. Phys. A Vol.7
 pag.6845(1992).
\bibitem{tp8} J.A.Wheeler, R.P.Feynman: Rev. Mod. Phys. A Vol.17,
 pag.157(1945).
\bibitem{tp9} C.G.Bollini, L.E.Oxman, M.C.Rocca: ``An equivalence
 theorem for higher order equations coupled to a gauge field''.
 La Plata preprint, December (1995).
\bibitem{tp10} L.D.Faddeev and A.A.Slavnov: ``Gauge Fields. Introduction
 to Quantum Theory''. The Benjamin-Cummings Publishing Company,
 Inc.(1970).
\bibitem{tp11} C.G.Bollini, L.E.Oxman: Int. J. of Mod. Phys. A Vol.8,
 N.18, pag.3185(1993).

\end{thebibliography}
\end{document}